\documentclass[prl,aps,twocolumn,floats,nofootinbib]{revtex4}
\usepackage{amsmath,amssymb,graphicx,psfrag}
\begin{document}
\renewcommand{\ni}{{\noindent}}
\newcommand{\dprime}{{\prime\prime}}
\newcommand{\be}{\begin{equation}}
\newcommand{\ee}{\end{equation}}
\newcommand{\bea}{\begin{eqnarray}} 
\newcommand{\eea}{\end{eqnarray}}
\newcommand{\la}{\langle}
\newcommand{\ra}{\rangle} 
\newcommand{\dg}{\dagger}
\newcommand\lbs{\left[}
\newcommand\rbs{\right]}
\newcommand\lbr{\left(}
\newcommand\rbr{\right)}
\newcommand\f{\frac}
\newcommand\e{\epsilon}
\newcommand\ua{\uparrow}
\newcommand\da{\downarrow}

\title{ Is there a superconducting phase in the half-filled ionic Hubbard model ?}
\author{Anwesha Chattopadhyay$^{1}$, Soumen Bag$^{2}$, H. R. Krishnamurthy$^{2}$, Arti Garg$^{1}$ }
\affiliation{$^{1}$ Condensed Matter Physics Division, Saha Institute of Nuclear Physics, HBNI, 1/AF Bidhannagar, Kolkata 700 064, India\\
$^{2}$ Department of Physics, Indian Institute of Science, Bangalore 560 012, India}
\begin{abstract}
We investigate the ionic Hubbard model (IHM) at half-filling in the limit of strong correlations and large ionic potential. The low energy effective Hamiltonian in this limit, obtained by a similarity transformation, is a modified $t-J$ model with effective second neighbour hopping terms.
We explore the possibilities of d-wave pairing and extended s-wave pairing superconducting (SC) phases on a two dimensional square lattice at zero temperature within a Gutzwiller projected renormalized mean field theory. In the sector of solutions that forbid spin ordering, the system shows a finite non-zero d-wave as well as extended s-wave pairing amplitude for $\Delta \sim U \gg t$. The width of the superconducting phase in $U-\Delta$ regime shrinks with increase in $U$ and $\Delta$, though the extended s-wave pairing phase is higher in energy than the d-wave pairing superconducting phase. But in a spin resolved renormalized mean field calculation, which allows for an antiferromagnetic (AF) order along with the d-wave or extended s-wave pairing, the SC phase is no longer viable and the system shows a direct transition from an AF ordered phase to a paramagnetic band insulator. Except for a thin sliver of a half-metallic AF phase close to the AF transition point, most of the AF ordered phase is a Mott insulator. We benchmarked the AF Mott insulator to band insulator transition within the Gutzwiller projected renormalized mean field theory against the dynamical mean field theory (DMFT) solved using continuous time quantum Monte-Carlo (CTQMC). Our work suggests that the ground state phase diagram of the IHM at half-filling in the limit of extreme correlations does not have any SC phase. The SC phase seen in the paramagnetic sector is a metastable phase, being higher in energy than the AF  Mott insulator phase.  
\end{abstract}
\maketitle
\section{Introduction}

Doping a strongly correlated  Mott insulator (MI) away from commensurate filling results in a superconducting phase~\cite{RMP_Lee} as known from high $T_c$ cuprates and the recently discovered superconductivity in magic angle twisted bilayer graphene~\cite{BLG}. The minimal model to describe this physics is the strongly correlated Hubbard model, which at half-filling maps onto an effective Heisenberg model having an AF  insulating ground state and doping holes or electrons into this system results in a superconducting state.

In this work, we study a variant of the Hubbard model, known as the ionic Hubbard model (IHM), which is basically the Hubbard model defined on a bipartite lattice with an additional staggered potential $\Delta$. The physics of IHM is governed by the competition between the staggered potential $\Delta$ and the Hubbard $U$~\cite{1d-1,1d-2,1d-3,AG1}. At half-filling, in the large $U$ limit, the system is a MI while for large $\Delta$ regime, the system is a band insulator (BI) due to doubling of the unit cell. The physics of the intermediate regime in which $U\sim \Delta$, straddling the two insulating phases, has been of interest to the condensed matter community.
In this work we focus on the limit when $U \sim \Delta$ but both are much larger than the hopping amplitude $t$, that is, $U\sim \Delta \gg t$ and explore the possibility of a superconducting phase in this limit of the IHM at half-filling. 

The IHM has been realized for ultracold fermions on an optical honeycomb lattice~\cite{IHM_expt}. Due to recent developments in layered materials and heterostructures, it is indeed possible to think of many scenarios where the IHM can be used as a minimal model to understand the qualitative physics. Some of these examples are graphene on h-BN substrate where due to the difference in energy of B and N sites, electrons in the graphene sheet also feel a staggered potential. Also for a bilayer graphene in the presence of a transverse electric field, a potential difference is induced between the two layers~\cite{Castro} which plays the role of the staggered potential. Interactions are inevitably present in all real materials. 

The IHM has been studied in various dimensions by a variety of numerical and analytical tools~\cite{1d-1,1d-2,1d-3,Soos,AG1,Jabben,cdmft_ihm,hartmann,kampf,Hoang,AG2,rajdeep,soumen,qmc_ihm1,qmc_ihm2}. In one-dimension~\cite{1d-1,1d-2,1d-3,Soos} it has been shown to have a spontaneously dimerized phase which separates the weakly coupled BI from the strong coupling MI. 
In higher dimensions ($d > 1$), this model has been mostly studied in the weak to intermediate coupling regime for $\Delta \sim t$ by many groups using the dynamical mean field theory (DMFT)~\cite{AG1,Jabben,hartmann,kampf,AG2,rajdeep,soumen}, determinantal quantum Monte carlo~\cite{qmc_ihm1,qmc_ihm2}, and coherent potential approximation~\cite{Hoang}.  The solution of the DMFT self consistent equations for intermediate strength of $U$ and $\Delta \sim t$, in the paramagnetic sector at half filling at zero temperature shows an intervening correlation induced metallic phase~\cite{AG1,hartmann,Hoang,qmc_ihm1,qmc_ihm2}. When one allows for spontaneous spin symmetry breaking the transition from paramagnetic BI to antiferromagnetic (AF) insulator preempts the formation of the para-metallic phase~\cite{kampf,cdmft_ihm}, except, as shown in a recent paper coauthored by two of us~\cite{AG2}  using DMFT with iterated perturbation theory (IPT) as the impurity solver, for a sliver of a half-metallic AF  phase. Upon doping the IHM in the intermediate coupling regime for $\Delta \sim t$, one gets a broad ferrimangetic half-metal phase~\cite{AG2} sandwiched between a weakly correlated PM metal for small $U$ and a strongly correlated metal for large $U$.
Recently the IHM was solved at half-filling within DMFT using continuous time Monte Carlo (CTQMC) as an impurity solver~\cite{rajdeep,soumen}. In the large $U$ limit $U \gg (\Delta, t)$ it maps onto an effective Heisenberg model with the spin-exchange coupling $\tilde{J}=t^2U/(U^2-\Delta^2)$~\cite{rajdeep,soumen}. At any finite $T$, for $\Delta \sim t$, as $U$ increases, first the magnetic order turns on via a first order phase transition followed up by a continuous transition back to the PM phase.  There is a line of tricritical point $T_{tcp}$ that separates the two surfaces of first and second order phase transitions~\cite{soumen}. 

In this paper we study the half-filled IHM in the limit where {\it{both}} the Hubbard $U$ and the staggered potential $\Delta$ are much larger than the hopping amplitude. Cluster DMFT study in this limit~\cite{cdmft_ihm} demonstrated a direct transition between the AF  MI and the BI as $\Delta$ is increased for a fixed large value of $U$. Recently this limit has been explored using slave-boson mean field theory~\cite{samanta} which demonstrated a transition from MI to BI as $\Delta$ increases followed up by a transition to a broad superconducting phase as $\Delta$ is increased further. Clearly there is no clear consensus on the phase diagram of the IHM in this limit. In order to develop some understanding of the IHM in this limit, here we solve it using a Gutzwiller projected renormalized mean field theory as well as using the DMFT+CTQMC technique. Below we summarize our main findings from this analysis. 

The IHM we study is on a 2-dimensional square lattice, at zero temperature. We find that within a spin symmetric Gutzwiller projected mean field theory, the d-wave pairing does indeed turn on for a small range of $\Delta \sim U$ sandwiched between a paramagnetic MI and a BI. Though the extended s-wave pairing amplitude is also non zero for a small $\Delta$ range, it is always a little higher in energy than the d-wave superconducting phase. But in a generic calculation, where the system is allowed to have phases with broken spin symmetry as well, the AF Mott insulating phase wins over the superconducting phase, and the system does not have any stable superconducting ground state. There occurs a transition from the AF MI to the paramagnetic BI, with a thin half-metallic phase intervening between the two insulators close to the transition point. This phase diagram shows consistency with the earlier analysis~\cite{AG1,soumen} in weak to intermediate $U$ and $\Delta$ regime, where a metallic phase is observed within a spin symmetric calculation; however, once spin-ordering is allowed for, the AF  MI preempts the formation of metal, except for a thin half-metallic phase close to the transition between the MI and the BI. Hence there is a continuity in the phase diagram along the $U \sim \Delta$ line as $U$ increases.  Surprisingly, the phase diagram obtained from the Gutzwiller projected mean field calculation differs from the one obtained from the slave boson mean field theory calculation~\cite{samanta} where a broad SC phase appears beyond the BI phase as $\Delta$ increases. We have benchmarked the AF transition point obtained within the Gutzwiller projected mean field theory calculation against the DMFT+CTQMC calculation which has earlier been shown to capture the correct strongly correlated limit of IHM~\cite{rajdeep,soumen} within a mean field description of the AF order.

The rest of this paper is organized as follows. In section II, we describe the model, the low energy Hilbert space which is relevant to the limit $U \sim \Delta \gg t$, and the effective low energy Hamiltonian , obtained using a similarity transformation. Furthermore, we describe the Gutzwiller approximation used to solve this low energy Hamiltonian. In section III, we briefly describe the Gutzwiller projected renormalized mean field theory (RMFT) for the AF phase and then benchmark our results against the DMFT+CTQMC calculations. In section IV, we describe the spin symmetric RMFT calculation which allows for superconducting pairing amplitude followed up by the generic RMFT calculation in section V where we include the pairing amplitude as well as the magnetic order. At the end we conclude and summarize.

\section{Model and Method: Low energy effective Hamiltonian and Gutzwiller Approximation}
The IHM is described on a bipartite lattice by the Hamiltonian,
\[
  H=-t\sum_{<i,j>,\sigma}(c_{i\sigma}^{\dagger}c_{j\sigma} + h.c.)-\f{\Delta}{2}\sum_{i \in A}n_i+\f{\Delta}{2}\sum_{i \in B} n_i \]
\be
\mbox{~~~~~~~~~~~~~~~~~~~~}+U\sum_{i}n_{i\uparrow}n_{i\downarrow}-\f{U}{2}\sum_{i}n_{i}
\label{model}
\ee
Here $t$ is the nearest neighbor hopping, $\Delta$ is the staggered one body potential and $U$ is the onsite Hubbard repulsion. At half-filling, corresponding to $(\la n_A\ra+\la n_B \ra)/2=1$, the Hamiltonian is particle-hole symmetric,  with $\mu=\f{U}{2}$.

In the limit $U\sim \Delta \gg t$, the $t=0$ model can be thought of as the unperturbed model and the hopping can be treated perturbatively. For $t=0$, and $U\sim \Delta$, from the energies associated with all possible configurations at each site, it is easy to see that holes on the A sublattice are energetically expensive and doublons are energetically unfavorable on the B sites. Hence holes on A and doublons on B sublattice get eliminated from the low energy Hilbert space. 
As shown in a work coauthored by two of us~\cite{Anwesha}, the effective low energy Hamiltonian in the limit $U \sim \Delta \gg  t$, obtained by a similarity transformation which eliminates processes which inter-connect the high and low energy sector of the Hilbert space is given by
\be
\mathcal{H}_{eff} = H_0+H_{t,low}+H_{d}+H_{tr}+H_{ex}
\label{Heff}
\ee
Here $H_{t,low}$ is the hopping process in the low energy Hilbert space.  As an effect of projection of holes and doublons from A and B sublattice, respectively, many of the nearest neighbour hopping processes between sites of sublattice A and B, where either the initial or the final state has holes on $A$ sublattice and doublons on $B$ sublattice, belong to the high energy sector of the Hilbert space and hence get projected out from low energy Hamiltonian. But interestingly, in the half filled IHM there are hopping processes which belong only to the low energy Hilbert space, e.g. $|d_A0_B\ra \Leftrightarrow |\ua_A\da_B\ra$. This is in contrast to the half-filled Hubbard model~\cite{Fazekas}, where hopping is completely projected out of the low energy Hilbert space. Hence we have the following expression for $H_{t,low}$
\bea
    H_{t,low}=-t\sum_{<ij>,\sigma}\tilde{c}_{iA\sigma}^{\dagger}\tilde{\tilde{c}}_{jB\sigma}+\tilde{\tilde{c}}_{jB\sigma}^{\dagger}\tilde{c}_{iA\sigma} \nonumber \\
= -t\sum_{<ij>,\sigma} \mathcal{P}[c^\dagger_{iA\sigma}c_{jB\sigma}+h.c.]\mathcal{P}
\label{order}
\eea

Here $P$ is the projection operator that projects out holes from sublattice A and doublons from sublattice B. The new fermionic operators in the projected Hilbert space are defined as 
\begin{equation}
        \tilde{c}_{A\sigma}^{\dagger}\equiv \eta (\sigma)X_{A}^{d \leftarrow \bar{\sigma}}=c_{A\sigma}^{\dagger}n_{A\bar{\sigma}}
\label{cnewA}
    \end{equation}
     \begin{equation}
          \tilde{\tilde{c}}_{B\sigma}^{\dagger} \equiv X_{B}^{\sigma \leftarrow 0}=c_{B\sigma}^{\dagger}(1-n_{B\bar{\sigma}})
\label{cnewB}
     \end{equation}

A second order hopping process starting from and returning to the sector of states with single occupancies on two neighbouring sites, where the first hopping results in a virtual hole on A and a doublon on B, results in an effective spin exchange process $H_{ex}$ given by
\[H_{ex}=-J_1\sum_{<ij>,\sigma}[\tilde{c}_{iA\bar{\sigma}}\tilde{c}_{iA\bar{\sigma}}^{\dagger}\tilde{\tilde{c}}_{jB\bar{\sigma}}^{\dagger}\tilde{\tilde{c}}_{jB\bar{\sigma}}-\tilde{c}_{iA\sigma}\tilde{c}_{iA\bar{\sigma}}^{\dagger}\tilde{\tilde{c}}_{jB\sigma}^{\dagger}\tilde{\tilde{c}}_{jB\bar{\sigma}}] \]
\vskip-0.4cm
\be
= 2J_{1}\sum_{<ij>}\mathcal{P}(S_{iA}.S_{jB}-(2-n_{iA})n_{jB}/4)\mathcal{P}
\ee

with $J_1=\frac{t^2}{U+\Delta}$. There are dimer processes $H_{d}$ where a spin from an A site hops to an empty B site, and then hops back to the same A site, creating a virtual state with a hole on the A site. In another dimer process, a spin $\sigma$ from a doubly occupied A site hops to a B site which has $\bar{\sigma}$ and then hops back to A site, resulting in a virtual state with a doublon on B site. Both these processes are of order $t^2/\Delta$ and can be written as 
\[H_{d}=-\frac{t^2}{\Delta}\sum_{\sigma,<ij>}\big[\tilde{c}_{iA\bar{\sigma}}\tilde{c}_{iA\bar{\sigma}}^{\dagger}\tilde{\tilde{c}}_{jB\sigma}\tilde{\tilde{c}}_{jB\sigma}^{\dagger}+\tilde{c}_{iA\sigma}^{\dagger}\tilde{c}_{iA\sigma}\tilde{\tilde{c}}^\dagger_{jB\bar{\sigma}}\tilde{\tilde{c}}_{jB\bar{\sigma}}\big]\]
\vskip-0.4cm
\be
=-\frac{t^2}{\Delta}\sum_{<ij>,\sigma}\mathcal{P}\big[(1-n_{iA\bar{\sigma}})(1-n_{jB})+(n_{iA}-1)n_{jB\bar{\sigma}}\big]\mathcal{P}
\label{hd_hopp}
\ee

 Trimer terms, leading to $H_{tr}$, correspond to the hopping of a doublon or a hole from a site on the $A(B)$ sublattice to it's second neighbour site in the same sublattice via a two hop process. Effectively, there is a doublon
 hopping which is intra A sublattice, where as the hole hopping is intra B sublattice. In terms of projected operators, these are represented as
  
\[
H_{tr}=-\frac{t^2}{\Delta}\sum_{\sigma,<ijk>}(\tilde{c}_{kA\sigma}^{\dagger}\tilde{\tilde{c}}_{jB\bar{\sigma}}^{\dagger}\tilde{\tilde{c}}_{jB\bar{\sigma}}\tilde{c}_{iA\sigma}+\tilde{c}_{iA\bar{\sigma}}\tilde{\tilde{c}}_{jB\bar{\sigma}}^{\dagger}\tilde{\tilde{c}}_{jB\sigma}\tilde{c}_{kA\sigma}^{\dagger})\]
\vskip-0.4cm
\[-\frac{t^2}{\Delta}\sum_{\sigma,<jil>}(\tilde{\tilde{c}}_{lB\sigma}\tilde{c}_{iA\bar{\sigma}}\tilde{c}_{iA\bar{\sigma}}^{\dagger}\tilde{\tilde{c}}_{jB\sigma}^{\dagger}+\tilde{\tilde{c}}_{jB\sigma}^{\dagger}\tilde{c}_{iA\sigma}\tilde{c}_{iA\bar{\sigma}}^{\dagger}\tilde{\tilde{c}}_{lB\bar{\sigma}})\]
\bea
   =-\frac{t^2}{\Delta}\sum_{\sigma,<ijk>}\mathcal{P}(c^{\dagger}_{kA\sigma}n_{jB\bar{\sigma}}c_{iA\sigma}+
    c_{iA\bar{\sigma}}c^{\dagger}_{jB\bar{\sigma}}c_{jB\sigma}c^{\dagger}_{kA\sigma})\mathcal{P}\nonumber \\
-\frac{t^2}{\Delta}\sum_{\sigma,<jil>}\mathcal{P}(c_{lB\sigma}[(1-n_{iA\bar{\sigma}})c_{jB\sigma}^{\dagger}+
     c_{iA\sigma}^{\dagger}c_{iA\bar{\sigma}}c_{jB\bar{\sigma}}^{\dagger}])\mathcal{P}
\label{intra_hop}
    \eea
The effective low energy Hamiltonian mentioned above can not be solved using regular perturbation theory because the projected fermionic operators $\tilde{c}_A$ and $\tilde{\tilde{c}}_B$ do not satisfy the standard anti-commutation relations of canonical fermions and hence Wick's theorem can not be applied. The possible approaches to solve $H_{eff}$ are either fully numerical, like variational Monte-Carlo (VMC)~\cite{VMC} where the projection constraints can be handled exactly in each configuration but is computationally very expensive, or one can use the Gutzwiller approximation in the same spirit as it is used for doublon projection in the $t-J$ model~\cite{Gutzwiller,GA,Mohit,AGnature}.  Within the Gutzwiller approximation, the effect of projection is treated approximately by renormalizing the coefficients of the various terms in $H_{eff}$ by corresponding Gutzwiller factors and calculating the expectation value of the renormalized Hamiltonian in the unprojected basis. 
  The Gutzwiller factors, for the half-filled IHM in the limit $U \sim \Delta \gg t$, for the hole projection from the A sublattice and the doublon projection from the B sublattice have been calculated in an earlier work coauthored by two of us~\cite{Anwesha}. The renormalized Hamiltonian obtained is of the form,
\[
\tilde{H}=H_0-t\sum_{\sigma,<ij>}g_{t\sigma}[{c}_{iA\sigma}^{\dagger}{c}_{jB\sigma}+{c}_{jB\sigma}^\dagger{c}_{iA\sigma}] \]
\vskip-0.5cm
\[
-g_{1}\frac{t^2}{\Delta}\sum_{<ij>,\sigma}[(1-n_{iA\bar{\sigma}})(1-n_{jB})+(n_{iA}-1)n_{jB\bar{\sigma}}] \]
\vskip-0.5cm
\[
-\frac{t^2}{\Delta}\sum_{\sigma,<ijk>}(g_{t\sigma}c^{\dagger}_{kA\sigma}n_{jB\bar{\sigma}}c_{iA\sigma}+g_2 c_{iA\bar{\sigma}}c^{\dagger}_{jB\bar{\sigma}}c_{jB\sigma}c^{\dagger}_{kA\sigma} )+ h.c.\]
\vskip-0.5cm
\[
-\frac{t^2}{\Delta}\sum_{\sigma,<jil>}(g_{t\sigma} c_{lB\sigma}(1-n_{iA\bar{\sigma}})c_{jB\sigma}^{\dagger}+g_2 c_{lB\sigma}c_{iA\sigma}^{\dagger}c_{iA\bar{\sigma}}c_{jB\bar{\sigma}}^{\dagger})\]
\vskip-0.5cm
\be
+J_1\sum_{<i,j>}[g_{s}S_{iA}.S_{jB}-\f{1}{4}(2-n_{iA})n_{jB}]
 \label{H_ren}
\ee

 Here $g_{t\sigma},g_{1},g_s$ and $g_2$ are the Gutzwiller renormalization factors.
 The factors for various processes in $H_{eff}$ were calculated under the approximation that the densities on A and B sites before and after the projection are the same. Table~\ref{GAtab} provides expressions for the various Gutzwiller factors in terms of the mean field quantities, namely, $\delta=(n_A-n_B)/2$,  the density difference between the two sublattices, and $m_s=(m_A-m_B)/2$, the staggered magnetization in the symmetry broken antiferromagnetic phase.

  \large
\begin{table}[h!]
\begin{center}
 \begin{tabular}{||c | c ||} 
 \hline
 Gutzwiller Factors & Expressions\\[2ex]
 
 \hline
 
$ g_{t\sigma}$ &  $\f{2\delta}{1+\delta+\sigma m_s}$ \\ \hline
 $g_{s}$  &  $\f{4}{(1+\delta)^2-m_s^2}$ \\ \hline
$ g_{1}$  & $1$ \\ \hline
$g_{2}$  & $\f{4\delta}{(1+\delta)^2-m_s^2}$ \\ [1ex] 
 \hline
\end{tabular}

\end{center}
\caption{Gutzwiller factors for various terms in $H_{eff}$ at half-filling in the antiferromnagnetically ordered phase~\cite{Anwesha}.}
\label{GAtab}
\end{table}
 \normalsize
Note that for $m_s=0$, the expressions for $g_{t}$ and $g_{s}$ become similar to that of the familiar hole-doped tJ model with $\delta$ in IHM playing the role of doping in tJ model~\cite{GA} although the projection constraints in the two situations are completely different.
  
 $H_{0}$, the unperturbed part of the Hamiltonian in the projected space is equivalent to $H_0=\sum\limits_{i}\f{U-\Delta}{2}[n_{iA\uparrow}n_{iA\downarrow}+(1-n_{iB\uparrow})(1-n_{iB\downarrow})]$.
To see this, consider first the A sublattice, where holes are not allowed in the low energy Hilbert space. The unperturbed Hamiltonian can be written as $H_{0,A}= \mathcal{P}_{h}\bigg[U(1-n_{A\uparrow})(1-n_{A\downarrow})+\bigg(\f{U-\Delta}{2}\bigg)n_{A}\bigg]\mathcal{P}_{h}$. Since holes are projected out, only the second term survives under the projection. Using the completeness relation in the hole projected Hilbert space, $n_{A\ua}(1-n_{A\da})+n_{A\da}(1-n_{A\ua})+n_{A\ua}n_{A\da}=1$, one can show that $\mathcal{P}_{h}n_A\mathcal{P}_{h} \equiv (1+n_{A\ua}n_{A\da})$.
 Similarly,  on the B sublattice where doublons are not energetically favourable $H_{0,B} = \mathcal{P}_{d}\bigg[Un_{B\uparrow}n_{B\downarrow}-\bigg(\f{U-\Delta}{2}\bigg)n_{B}\bigg]\mathcal{P}_{d}$ where only the second term survives. Using the completeness relation on the B sublattice, $H_{0,B} =(U-\Delta)/2[(1-n_{B\ua})(1-n_{B\da})-1]$.

We have solved the renormalized low energy effective Hamiltonian within a mean field theory. Before we go into details of this renormalized mean field theory (RMFT) and the phase diagram obtained from it, below we first benchmark the results obtained from RMFT against DMFT+CTQMC. \\
\\

\section{Benchmarking the renormalized Hamiltonian and Gutzwiller Approximation}
The Gutzwiller approximation for the projection of doublons done for the hole-doped $t$-$J$ model has shown qualitative and quantitative consistency with results obtained from VMC~\cite{Mohit}. Hence we expect that the Gutzwiller approximation for the projection of holes and doublons from A and B sublattice sites, respectively, will also capture the physics qualitatively correctly. To check the validity of this expectation, in this  section we compare the results obtained within RMFT against those obtained from DMFT+CTQMC. DMFT+CTQMC has been shown to capture the physics of strong correlations and the projection correctly in the limit $U \gg \Delta,t$ as demonstrated by the correct dependence of Neel temperature for the AF  order as a function of $\Delta$~\cite{soumen,rajdeep}. 

However, within a single site DMFT, we can not explore the possibility of d-wave or extended s-wave superconductivity. Hence our comparison of the results of RMFT with the DMFT+CTQMC calculations is without including the superconducting pairing amplitude as a mean field. To be precise, we give non zero expectation values only to (a) the staggered magnetization $m_\alpha=\la c_{i\alpha\ua}^\dagger c_{i\alpha\ua}-c_{i\alpha\da}^\dagger c_{i\alpha\da}\ra$, (b) the density difference between two sublattices $\delta = \la (n_A-n_B)\ra /2$,  (c) the inter-sublattice fock shift $\chi_{\sigma AB} = \la c^\dagger_{i\sigma A}c_{j\sigma B}\ra $, and (d) the intra-sublattice fock shifts $\chi_{\alpha\alpha}=\la c_{i\sigma\alpha}^\dagger c_{j\sigma\alpha}+h.c. \ra$. Here $\alpha$ is the sublattice index and $\sigma$ is the spin index.  
 The mean field quadratic Hamiltonian can be written as 
\bea
H_{MF} = \sum_{k,\sigma}h_{1\sigma}(k)[c_{kA\sigma}^\dagger c_{kA\sigma}-c_{kB\sigma}^\dagger c_{kB\sigma}] \nonumber \\
+h_{2\sigma}(k)[c_{kA\sigma}^\dagger c_{kB\sigma}+h.c.]
\eea 
 where 
\begin{widetext}
 $h_{1\sigma}(k)=\f{U-\Delta}{2}\bigg(\f{1+\delta-\sigma m}{2}\bigg)-\f{t^2}{\Delta}\bigg[4(1-2\delta)+g_{t\bar{\sigma}}(2\chi_{BB\bar{\sigma}}+4\chi_{BBxy\bar{\sigma}})+g_{t\sigma}\f{1-\delta+\sigma m}{2}\gamma_{k}^{'}\bigg]-\f{2t^2}{U+\Delta}g_{s}\sigma m+\f{2t^2}{U+\Delta}(1-\delta)$

$h_{2\sigma}(k)=\bigg[-tg_{t\sigma}-\f{t^2}{\Delta}(-2\chi_{AB\sigma}+6g_{2}\chi_{AB\bar{\sigma}})-\f{t^2}{U+\Delta}[g_s(\f{1}{2}\chi_{AB\sigma}+\chi_{AB\bar{\sigma}})+\f{1}{2}\chi_{AB\sigma}]\bigg]\gamma_{k}$
\be
\ee
Here, $\gamma_{k}=2[\cos{(k_{x})}+\cos{(k_{y})}]$ and $\gamma_{k}^{'}=$$2[\cos{(2k_{x})}+\cos{(2k_{y})}]+4[\cos{(k_{x}+k_{y})}+\cos{(k_{x}-k_{y})}]$.
\end{widetext}

The mean field Hamiltonian $H_{MF}$ can be diagonalized using standard canonical transformation $c_{kA\sigma}=\alpha_{k\sigma}d_{k1\sigma}+\beta_{k\sigma}d_{k2\sigma}$ and $c_{kB\sigma}=\alpha_{k\sigma}d_{k2\sigma}-\beta_{k\sigma}d_{k1\sigma}$ where $\alpha$ and $\beta$ are fixed such that the off-diagonal part of Hamiltonian written in terms of the $d$ operators vanishes. This results in $2\alpha_{k\sigma}^{2}=(1-h_{1\sigma}(k)/E_\sigma(k))$  and $2\beta_{k\sigma}^{2}=(1+h_{1\sigma}(k)/E_\sigma(k))$ with $E_\sigma(k)=\sqrt{h_{1\sigma}(k)^{2}+h_{2\sigma}(k)^{2}}$.

\begin{figure}[!htbp]
    \centering
    \includegraphics[width=5.2cm,angle=-90]{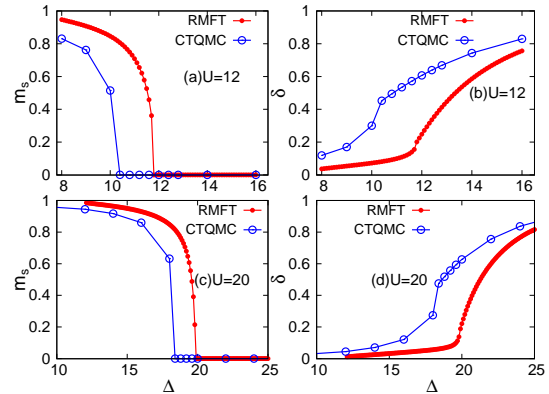}
    \caption{Staggered magnetization, $m_s$ and the density difference $\delta$ vs $\Delta$ for $U=12t$ and $20t$. Blue circles show the data obtained in a DMFT+CTQMC calculation and the red data points are obtained within a Gutzwiller projected RMFT calculation respectively.}
\label{mf1}
  \end{figure}

At half filling, the magnetization on A and B sublattices are equal and opposite to each other owing to the particle-hole symmetry. Hence $m_s=(m_A-m_B)/2=m_A$. Self-consistent equations for various mean field order parameters are
\bea
m_s=&\langle n_{iA\uparrow}\rangle-\langle n_{iA\downarrow}\rangle =\f{1}{N}\sum_{k}(\alpha_{k\uparrow}^{2}-\alpha_{k\downarrow}^{2}) \nonumber\\
\delta=&\f{1}{2N}\sum_{k\sigma}(\alpha_{k\sigma}^{2}-\beta_{k\sigma}^{2}) \nonumber\\
\chi_{AB\sigma}&=-\f{1}{4N}\sum_{k}\gamma_{k}\alpha_{k\sigma}\beta_{k\sigma}\nonumber\\
 \chi_{BB\sigma}&=\f{1}{N}\sum_{k}[\cos{(2k_{x})}+\cos{(2k_{y})}]\beta_{k\sigma}^{2}\nonumber\\
\chi_{BBxy\sigma}&=\f{1}{N}\sum_k 2 \beta^2_{k\sigma}\cos{(k_{x})}\cos{(k_{y})}
\label{sc_mf1}
\eea
The DMFT is done using CTMQC as an impurity solver using the hybridization expansion method, details of which can be found in our earlier work~\cite{soumen}. 
  Below we compare the staggered magnetization and the density difference obtained from the RMFT at $T=0$ for a half-filled IHM on the 2d-square lattice with those obtained from the DMFT+CTQMC at $\beta=50/t$ where $\beta$ is the inverse temperature. Fig.~\ref{mf1} shows qualitative as well as quantitative consistency between the Gutzwiller projected RMFT and the DMFT+CTQMC calculations for $U=12t$ and $U=20t$. In contrast, in slave boson mean field calculations~\cite{samanta} in the same limit one obtains the staggered magnetization transition point at $\sim 15.8t$ for $U=20t$ and also the value of $m_s$ is much smaller as compared to what is obtained within the RMFT or the DMFT+CTQMC calculations. The transition in both the calculations is first order, as reflected  in the jump in the magnetization at the transition point. Furthermore, the consistency between the RMFT and DMFT+CTQMC calculations improves for larger values of $U$ and $\Delta$, as expected.

\begin{figure}[!htbp]
\centering
\includegraphics[width=3.2cm,angle=-90]{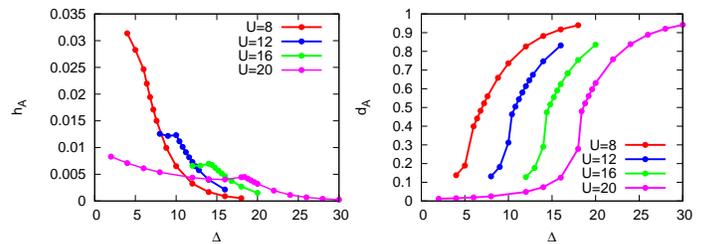}
\caption{Hole occupancy and double occupancy on A sites as a function of $\Delta$ obtained from the DMFT+CTQMC calculation for the IHM at half-filling on a  2-d square lattice.}
\label{nh_nd}
\end{figure}

We have also calculated the density of holes $h_A=\la (1-n_{\ua A})(1-n_{\da A})\ra$ and doublons $d_A=\la n_{\ua A}n_{\da A} \ra$ on A sublattice within DMFT+CTQMC.  Due to the p-h symmetry at half-filling, $h_A=\la n_{\ua B}n_{\da B}\ra = d_B$ and $h_B=d_A$. Fg.~\ref{nh_nd} shows the density of holes and doublons on the A sublattice. As shown, sublattice $A$ has negligible fraction of holes for $U\sim \Delta \ge 12t$. The density of holes decreases as $U$ increases and also for a fixed $U\ge 8t$, as $\Delta$ increases $h_A$ decreases becoming eventually less than one percent. This explains why a better consistency is observed at higher values of  $U$ and $\Delta$ between the DMFT+CTQMC calculation and the Gutzwiller projected RMFT theory, where holes from A sublattice and doublons from B sublattice have been fully projected out in the process of obtaining the low energy Hamiltonian.
\\
\section{Phase Diagram within Renormalised Mean Field Theory}
In this section, we provide details of two versions of the Gutzwiller projected RMFT calculations for the low energy Hamiltonian in Eq.~\ref{H_ren} allowing for the presence of a superconducting order parameter. One is the spin symmetric calculation where we do allow for a d-wave (or extended s-wave) pairing amplitude to have non-zero expectation value but $n_{\alpha \ua}= n_{\alpha\da}$ is imposed. The other is a less restricted calculation where we allow for superconductivity as well as symmetry breaking in the spin sector. 

Our solution of the mean field Hamiltonian involves a two step transformation. The Hamiltonian obtained after the first step of the transformation has both inter-band and intra-band pairing terms.  The results presented below are obtained by ignoring the inter-band pairing term, as it is smaller than the gap between the two bands at most of the points in the Brillouin zone, whence the second step of the transformation can be done analytically. Details of these calculations are given in Appendix A. In Appendix B, we have shown a comparison of these results with the calculations where the inter-band pairing term is kept, in which case the mean field Hamiltonian needs to be diagonalized numerically. As shown in Appendix B, at zero temperature, the contribution of the inter-band pairing term is negligible for most of the physical quantities of interest. Hence to obtain the zero temperature phase diagram it is a reasonably good approximation to ignore the inter-band pairing terms. \\

\subsection*{Results from Spin-symmetric RMFT}
In the spin symmetric RMFT, along with the mean fields mentioned earlier, we allow for a non zero value of the superconducting pairing amplitude $\Delta_{AB}(i,j) = \langle {c}_{iA\ua}^\dagger {c}_{j B\da}^\dagger -{c}_{i A\da}^\dagger {c}_{j B\ua}^\dagger\rangle $ looking for d-wave and extended s-wave pairing in the $U\sim \Delta \gg t$ limit of the half-filled IHM on a 2d square lattice. For d-wave pairing $\Delta_{AB}(i,i\pm x) = \Delta_d = -\Delta_{AB}(i,i\pm y)$ while for the extended s-wave $\Delta_{AB}(i,i\pm x)=\Delta_{AB}(i,i\pm y) = \Delta_s$. This implies $\Delta_{AB}(k) = 2\Delta_{d}[cos(k_x)-cos(k_y)]$ for the d-wave pairing while for the extended s-wave $\Delta_{AB}(k)=2\Delta_{s}[cos(k_x)+cos(k_y)]$.
We impose the spin symmetry $\la n_{i \ua} \ra =\la n_{i \da}\ra$, which further implies that all the inter- sublattice and intra sublattice Fock shifts are spin independent. Details of the mean-field calculations are given in Appendix A.
  
 \begin{figure}[!htbp]
    \centering
    \includegraphics[width=5cm,angle=-90]{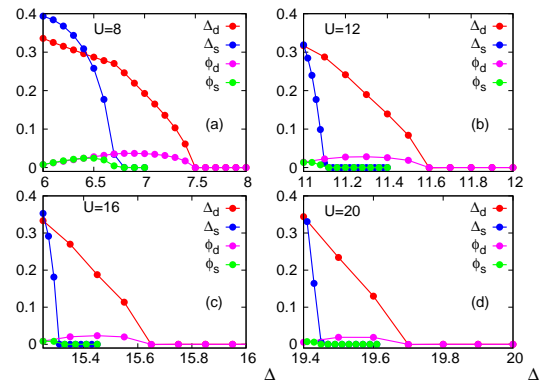}
    \caption{The superconducting pairing amplitude for d-wave and extended s-wave symmetry vs $\Delta$ obtained from spin symmetric RMFT. The pink curves shows the d-wave order parameter $\phi_d$ vs $\Delta$ while the green data points represent the extended s-wave order parameter $\phi_s$. Different panels show results for different values of $U$ ranging from $U=8t$ to $U=20t$. The extended s-wave pairing is observed for a smaller $\Delta$ regime while there is non-zero d-wave pairing amplitude for a comparatively broader range of $\Delta$.}
\label{PA_s}
  \end{figure}
  Fig.~\ref{PA_s} shows the pairing amplitude with the d-wave and the extended s-wave symmetry as a function of $\Delta$ for four values of $U$. Both the pairing amplitudes are non-zero for a finite range of $\Delta$ close to but less than $U$. For most of $U$ values of interest, the range of $\Delta$ over which the extended s-wave pairing appears is much smaller than the $\Delta$ range over which the d-wave pairing amplitude is non-zero. Note that though the pairing amplitude $\Delta_{d,s}$ remains non zero for values of $\Delta$ smaller than the range shown in Fig.~\ref{PA_s}, the density difference $\delta$ becomes close to zero for these smaller values of $\Delta$. This, as shown below, results in a vanishing SC order parameter for these smaller values of $\Delta$.  
 \begin{figure}[!htbp]
    \centering
    \includegraphics[width=5cm,angle=-90]{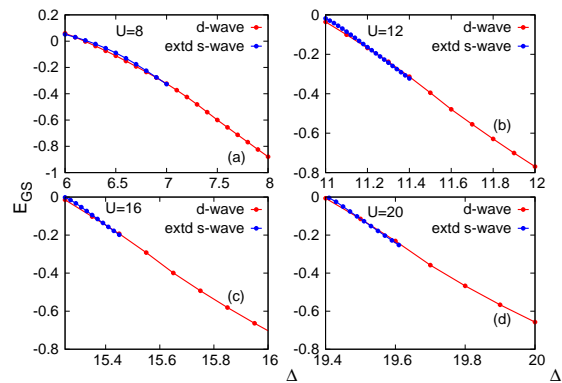}
    \caption{Ground state energy $E_{GS}$ vs $\Delta$ for extended s-wave pairing and d-wave pairing. For $\Delta$ ranges where the extended s-wave pairing amplitude is non zero, the ground state energy for the extended s-wave solution is higher than the ground state energy for the d-wave pairing superconducting phase.}
\label{Eg_ds}
  \end{figure}

Fig.~\ref{Eg_ds} further shows the comparison of the ground state energies for the self-consistent solutions with d-wave pairing and extended s-wave pairing. For almost the entire $\Delta$ regime where extended s-wave superconductivity is seen, the ground state energy of the extended s-wave superconducting phase is higher than that of the d-wave superconducting phase, making the latter the stable phase in the spin symmetric calculation.

\begin{figure}[!htbp]
    \centering
\vskip0.5cm
    \includegraphics[width=3cm,angle=-90]{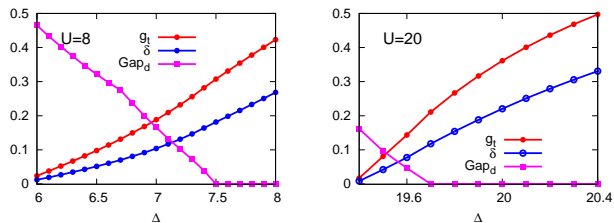}
\caption{Gutzwiller renormalization factor $g_t$ and the density difference $\delta$ vs $\Delta$ for the d-wave pairing SC phase. With increase in $\Delta$, the density difference between two sublattices increases, which results in enhanced coherent hopping of singlets. $Gap_d$ is the anti-nodal gap for the d-wave SC which, in contrast to the SC order parameter $\phi_d$, decreases monotonically with increase in $\Delta$.}
\label{gt}
\end{figure}

The superconducting order parameter $\phi_{d}$ and $\phi_s$ for the d-wave and extended s-wave channel respectively is defined as $\phi_{d,s}^{2}=g_{t}^{2}\lim_{r\rightarrow \infty}\langle c_{i\uparrow}^{\dagger}c_{j\downarrow}^{\dagger}c_{i+r\uparrow}c_{j+r\downarrow}\rangle$. For a given $U$, though the pairing amplitude is larger for smaller values of $\Delta$, because probability for formation of a singlet is larger for smaller $\Delta$, these singlets can hop around coherently only when there are sufficient number of doublons on $A$ sublattice and holes on $B$ sublattice. This can happen only when $n_A$ is sufficiently larger than and $n_B$ is sufficiently smaller than the average density of one. This is exactly what is indicated in the definition of the SC order parameter $\phi_{d,s}$ where $g_{t}$ is the Gutzwiller renormalization parameter for the kinetic energy. Fig.~\ref{gt} shows the behaviour of Gutzwiller factor $g_t$ as a function of $\Delta$ for d-wave pairing SC. For a given $U$, the density difference $\delta$ between two sublattices increases with increase in $\Delta$. This enhances the hopping between two sublattices through increase of $g_t$. On the other hand, the pairing amplitude $\Delta_{AB}$ decreases with increase in $\Delta$, resulting in a dome shaped non monotonic behaviour of $\phi$ as a function of $\Delta$ as shown in Fig.~\ref{PA_s}. 

Fig.~\ref{gt} also shows the antinodal gap $Gap_d=h_3(0,\pi)$ for the d-wave SC,  which is also the energy scale at which coherence peaks appear in the single particle density of states. Here $h_3(k)$ is the off-diagonal part of the mean-field Hamiltonian as shown in Appendix A. The antinodal gap monotonically decreases with increase in $\Delta$ as both the pairing amplitude $\Delta_d$ and the dominating Gutzwiller factor $g_s$ involved in $h_3(k)$ are monotonically decreasing functions of $\Delta$. 

The superconducting phase is sandwiched between two insulating phases. For $\Delta < \Delta_1$, where the SC order-parameter $\phi$ becomes non-zero first, the system is a paramagnetic MI with the gap in the single particle spectrum increasing monotonically with $U$. SC survives for $\Delta_1 < \Delta < \Delta_2$, and for $\Delta > \Delta_2$ the system goes into a trivial BI phase. The range in $\Delta$ for which the system shows the SC phase decreases with inreases in $U$. 
Note that the range of $\Delta$ for which the system shows the SC phase in this spin symmetric RMFT is much smaller than what is obtained using SBMFT~\cite{samanta}. 
 \subsection*{Results from Spin-asymmetric RMFT}
In the last section we showed that the half-filled IHM in the limit $U\sim \Delta \gg t$  has a d-wave superconducting phase on a 2d square lattice, provided the system is constrained to have spin symmetry. In this section, we carry out a less restricted calculation allowing for symmetry breaking in the spin sector as well and explore the fate of the SC phase in competition with the magnetic order in the system. Thus we give non zero values to the AF  order $m_s$ as well as to the superconducting pairing amplitude $\Delta_{AB}$ along with other mean fields like $\delta$ and the Fock shifts. The mean field Hamiltonian is then a $4 \times 4$ matrix for each allowed momentum $\vec{k}$ and requires a canonical transformation followed up by a Bogoliubov transformation to diagonalize it. Details of the mean field Hamiltonian, the transformations and the self-consistent equations for various order parameters are given in Appendix A and B. 

\begin{figure*}[!htbp]
    \centering
    \includegraphics[width=5cm,angle=-90]{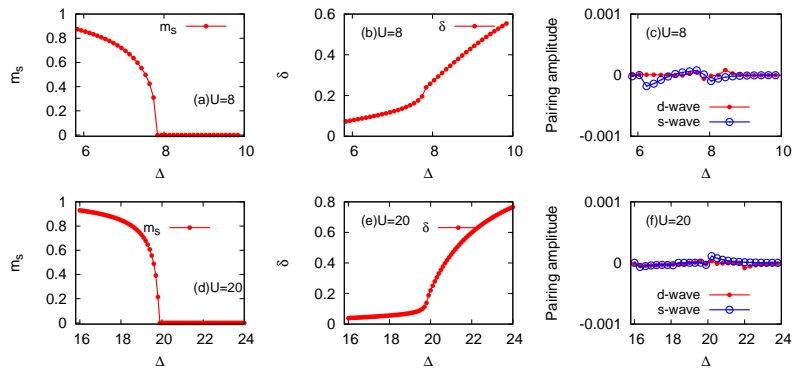}
    \caption {The staggered magnetization $m_s$, the density difference $\delta$ between the two sublattices and the d-wave and extended s-wave pairing amplitudes from the spin asymmetric calculation vs $\Delta$. The pairing amplitude remains vanishingly small for both the symmetries considered. Thus, the AF  order is energetically more stable than the SC order in the spin-asymmetric calculation.}
\label{sc_AF }
    \end{figure*}

 Fig.~\ref{sc_AF } shows the staggered magnetization $m_s$, the density difference between the two sublattices $\delta$ and the pairing amplitude with d-wave and extended s-wave symmetry for $U=8t$ and $U=20t$. Comparing with Fig.~\ref{PA_s}, we see that, for a fixed U, as $\Delta$ decreases from a large value the development of AF order preempts the formation of SC order, and hence the SC does not appear either with d-wave or extended s-wave symmetry. The system undergoes a direct transition from an AF MI into a paramagnetic insulator with possibility of only a thin half-metallic phase near the transition point, which we will discuss in a little while. 
Thus, though the recent SBMFT treatment of the half-filled IHM for $U \sim \Delta \gg t$ showed  a broad SC phase, our Gutzwiller projected RMFT suggests that the system has only a metastable d-wave SC phase, which is hidden under the AF  ordered phase. The SC phase is likely to get stablised only if the AF  order is frustrated somehow.

  \begin{figure}[!htbp]
    \centering
    \includegraphics[width=5cm,angle=-90]{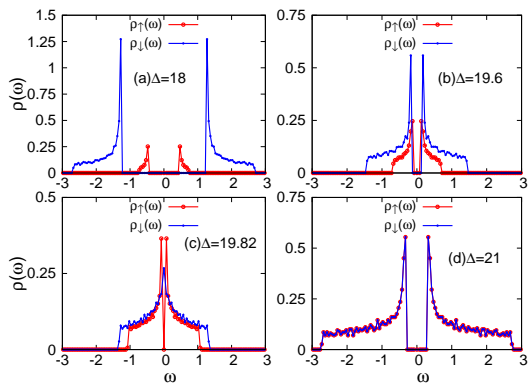}
    \caption{The single particle DOS for $U=20t$. (a) At $\Delta=18t$, the system has spin asymmetry with $gap_{\downarrow}>gap_{\uparrow}$. (b) Very near to $\Delta=19.6t$, the gaps are equal in both the spin channels but $\rho_{\uparrow} \neq \rho_{\downarrow}$.(c) At $\Delta=19.82t$, the system is a half-metal with down spin electrons conducting and up spin electrons insulating. (d) At $\Delta=21t$, the gaps are spin-symmetric with $\rho_{\uparrow}=\rho_{\downarrow}$. }
\label{dos}
  \end{figure}
Fig.~\ref{dos} shows the average single particle density of states (DOS) $\rho_\sigma(\omega) = 1/2\sum_\alpha \rho_{\alpha\sigma}(\omega)$. The spin-resolved sublattice single particle DOS is defined as

$\rho_{\alpha\sigma}(\omega)=-\dfrac{1}{\pi}\sum_{k}\text{Im }G_{\alpha\sigma}(k,\omega^{+})$

where, $\alpha$ represents the sublattice A or B and $\sigma$ is the spin index. Note that the Green's function in the projected Hilbert space is related to the Green's function $G_{\alpha\sigma}^{0}(k,\omega)$ in the unprojected space with appropriate Gutzwiller factor such that $G_{\alpha\sigma}(k,\omega)=g_{t\sigma}G_{\alpha\sigma}^{0}(k,\omega)$~\cite{AGnature}. As shown in Fig.~\ref{dos}, for $\Delta=18t$, $\rho_\sigma(\omega)$ is spin asymmetric with the gap in the down spin DOS being more than that in the up spin DOS. As we increase $\Delta$, the gaps in both channels as well as the asymmetry in the DOS for up and down spin channels decrease. Finally, at a particular $\Delta$ the gaps in both the channels become equal to each other, even though $\rho_{\uparrow} \neq \rho_{\downarrow}$, as is suggested by panel (b) of Fig.~\ref{dos}. After this the asymmetry in the up and down spin channel opens up again but now the gap in the up spin channel is more than that in the down spin channel [see panel (d) of Fig. 8]. As shown in panel (c) of Fig.~\ref{dos}, there is a sliver of $\Delta$ for which $\rho_\da(\omega=0)$ is non-zero indicating the metallic behaviour of the down-spin electrons while $\rho_\ua(\omega=0)$ is still zero with a small gap around $\omega=0$. This is the half-metallic point.
With a further finite increment in $\Delta$ the system makes a transition at $\Delta = \Delta_c$ to the band insulating phase with full spin symmetry in the DOS.

\begin{figure}[!htbp]
    \centering
    \includegraphics[width=5cm,angle=-90]{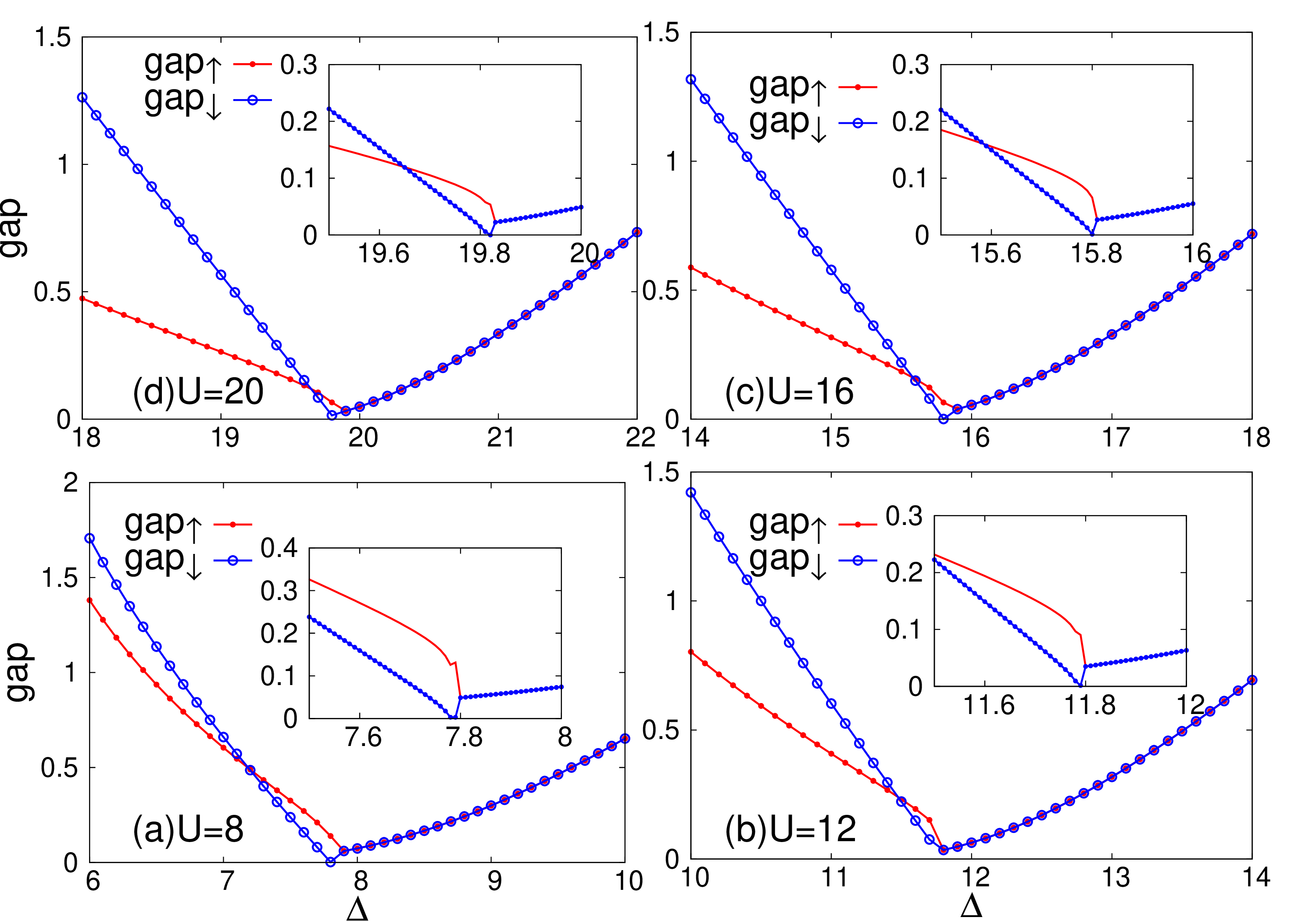}
    \caption{The gap in the single particle excitation spectrum for the up and down spin channels. For small $\Delta$, where the system has AF  order, $gap_{\downarrow}>gap_{\uparrow}$. On increasing $\Delta$, the gaps become equal and after that, $gap_{\uparrow} > gap_{\downarrow}$. Inset shows existence of a half-metallic state where $gap_{\downarrow}=0$. On further increase in $\Delta$, there is a transition to the paramagnetic BI phase, where the gaps are equal for the two spin components and increase with $\Delta$.}
\label{gap}
  \end{figure}
 Fig.~\ref{gap} shows that this behavior of the gaps in the single particle excitation spectrum for the up and down spin channels is similar for various values of $U$. For $\Delta<U$, the gaps are spin-asymmetric with the gap in the down spin channel being more than that in the up spin channel until at some $\Delta \stackrel{~}{<}\Delta_c$, the gaps cross and become equal. Post this crossing point, for $\Delta$ still below the transition point $\Delta_c$,  the gap in the up spin sector is more than that in the down spin sector. There occurs a point where gap in the down spin channel diminishes to zero (less than $0.001$ within our numerical calculations of the self-consistent mean field equations),  where as there is a finite gap in the up spin channel as shown in the inset. This indicates a half-metallic point within the AF phase but close to the transition into the BI phase. After the transition, for $\Delta > \Delta_c$, the system is in the spin-symmetric band-insulating phase where $gap_{\uparrow}=gap_{\downarrow}$.
 \begin{figure}[!htbp]
    \centering
    \includegraphics[width=5cm,angle=-90]{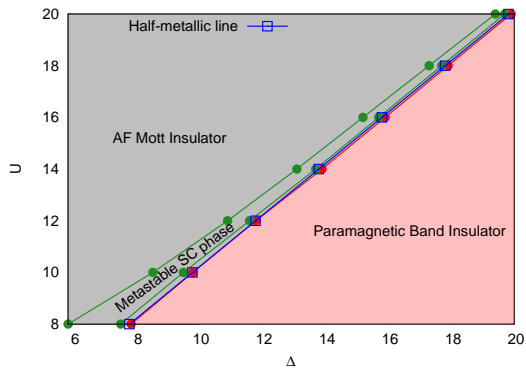}
    \caption{Complete phase diagram of the IHM in the $U\sim \Delta \gg t$  limit at half-filling on a  two dimensional square lattice, obtained within the Gutzwiller projected RMFT analysis. The system shows only one first order transition from an AF  ordered phase to a paramagnetic insulating phase. Most of the AF  ordered phase is a MI. Inside the AF phase, there is a metastable d-wave SC phase. Very close to the transition line between AF and the paramagnetic BI, the system shows a line of AF ordered half-metallic phase.}
\label{phase}
  \end{figure}

 Fig.~\ref{phase} shows the complete phase diagram of the IHM at half-filling in the $U\sim\Delta \gg t$ limit on a 2d square lattice obtained within the Gutzwiller projected RMFT. The system undergoes a first order transition from an AF  ordered state into the paramagnetic BI phase which is shown by the red line. Most of the AF phase is also Mott insulating in nature except for the thin half-metallic sliver close to the transition line, inside the AF  phase. Therefore, at the parameter values along this sliver there will be spin polarized conductivity in the system at half-filling. Inside the AF  phase, over the limited region shown, there also exists a metastable d-wave SC phase though the AF order is stabler than the SC order. Therefore, there is no stable superconducting phase in the IHM at half-filling in $U\sim\Delta \gg t$ regime within the Gutzwiller projected RMFT. This is in contrast to~\cite{samanta} where a robust extended s-wave SC phase is obtained within slave boson mean field theory.

The phase diagram we have obtained here using Gutzwiller projected RMFT in the limit $U\sim \Delta \gg t$ is adiabatically connected to the phase diagram obtained within DMFT (solved using CTQMC and iterative perturbative theory (IPT) as an impurity solver), for intermediate ranges of $U$ and $\Delta$~\cite{soumen}, where also a direct transition between AF  MI and the paramagnetic insulator is obtained except for a sliver of half-metallic phase. It is also consistent with the phase diagram obtained from cluster DMFT ~\cite{cdmft_ihm} where results were shown upto large values of $U$ and $\Delta$ and a direct transition between the MI and the paramagnetic BI is obtained. 
\section{Conclusions} 
In summary, in this paper we have studied the IHM at half-filling in the limit $U \sim \Delta \gg t$. The low energy effective Hamiltonian in this limit is defined on a projected Hilbert space where holes are projected out from one sublattice and the doublons are projected out from the other sublattice. Since the projected fermionic operators on either sublattice do not satisfy the algebra of canonical fermions, Wick's theorem does not hold for these operators and hence the effective low energy Hamiltonian can not be solved using standard perturbation theory. We implemented the Gutzwiller projection approximately by renormalizing the coefficients of the various terms in the effective Hamiltonian and solved the renormalized Hamiltonian within a mean field theory. On a 2d square lattice, we showed that the system has a d-wave superconducting phase sandwiched between a paramagnetic MI and a BI, provided the spin symmetry is enforced. But in a more general RMFT where the spin symmetry breaking is allowed, the AF order wins over the d-wave superconductivity. The system undergoes a transition from an AF MI to a paramagnetic BI with a thin sliver of a half-metallic phase in between, inside the AF Insulating region.

 It is surprising that though the Gutzwiller projected RMFT finds only a metastable SC phase, that too over a limited regime
in the $\Delta-U$ plane, slave boson mean field theory (SBMFT) on the other hand shows a broad stable SC region~\cite{samanta}. The RMFT treatment of the IHM gives AF order and the AF transition point which show consistency, both qualitatively and quantitatively, with the results obtained within DMFT+CTQMC; and the latter has been earlier shown to capture the correct physics of strong correlations and Gutzwiller projection in the limit $U \gg \Delta,t$~\cite{soumen,rajdeep}. Hence we expect that our RMFT results yield the correct strong correlation physics in the limit $U \sim \Delta \gg t$. Furthermore our study based on Gutzwiller projected RMFT is consistent with CDMFT study of IHM~\cite{cdmft_ihm}. Also the phase diagram within the RMFT is adiabatically continuous with the phase diagram obtained within DMFT (using IPT as well as CTQMC as impurity solver) for the weak to intermediate values of $U$ and $\Delta$~\cite{soumen}.

It will be interesting to explore the possibility of the explicit addition of a term to the IHM which can frustrate the AF  order and can stablise the SC phase. We hope to do this in future work. 
Also IHM has recently been implemented in the context of ultracold atoms~\cite{IHM_expt} where the relative strengths of $U$ and $\Delta$ can be tuned controllably. It will be really interesting to study this system in the limit $U \sim \Delta \gg t$ and to look for the superconducting phase experimentally.
\section*{Acknowledgements}
We would like to acknowledge Rajdeep Sensarma for discussions. 
\section{Appendix A}
In this Appendix, we provide details of the renormalized mean field theory where both, the SC order and the magnetic order, are allowed. We diagonalise the mean field Hamiltonian using a two step transformation. After the first step of the transformation, the effective Hamiltonian obtained has both inter-band and intra-band pairing terms. The inter-band pairing terms are much smaller then the gap between the two bands for most of the points on the Brilluion zone and should not contribute significantly at zero temperature. Hence we ignore the inter-band pairing terms which allows us to carry out the second step of the transformation also analytically. Below, we provide details of these transformations and the self consistent equations obtained for various order parameters. 
We also give results for the inter-sublattice and intra-sublattice Fock-shifts calculated within this mean field theory which were not presented in the section on results.
\\
{\bf Details of the renormalized mean field theory}:
The mean field quadratic Hamiltonian, where we have allowed for nearest neighbour spin-singlet pairing as well as spin ordering, is as follows,

\begin{widetext}
\begin{equation}
\mathcal{H} = \sum_{k}
\left({\begin{array}{cccc} c_{kA\uparrow}^{\dagger} &c_{-kA\downarrow}& c_{kB\uparrow}^{\dagger} & c_{-kB\downarrow} \end{array}}\right)
\left(\begin{array}{cccc} h_{1\uparrow}(k) & 0 &h_{2\uparrow}(k) & -h_3(k)\\ 0& -h_{1\downarrow}(k) & -h_3(k) & -h_{2\downarrow}(k)\\h_{2\uparrow}(k)&-h_3(k)& -h_{1\uparrow}(k)& 0\\-h_3(k)&-h_{2\downarrow}(k)&0&h_{1\downarrow}(k) \end{array}\right)
\left(\begin{array}{c} c_{kA\uparrow} \\c_{-kA\downarrow}^{\dagger}\\ c_{kB\uparrow} \\ c_{-kB\downarrow}^{\dagger} \end{array}\right) 
\end{equation}
\end{widetext}
The expressions for $h_{1\sigma}(k)$ and $h_{2\sigma}(k)$ are the same as given in section III.
For the d-wave symmetry the expression for $h_3(k)$ is 
$h_3(k)=\bigg[\dfrac{4t^2}{\Delta}(1-g_2)+\dfrac{4t^2}{U+\Delta}\bigg(\dfrac{3g_{s}}{4}-\dfrac{1}{4}\bigg)-\dfrac{2t^2}{\Delta}(g_{t\downarrow}+g_{t\ua})\bigg]\dfrac{\Delta_{AB}}{2}[\cos{(kx)}-\cos{(ky)}]$.
For the extended s-wave symmetry the expression is
$h_3(k)=\bigg[\dfrac{4t^2}{\Delta}(1+3g_2)+\dfrac{4t^2}{U+\Delta}\bigg(\dfrac{3g_{s}}{4}-\dfrac{1}{4}\bigg)+\dfrac{6t^2}{\Delta}(g_{t\downarrow}+g_{t\ua})\bigg]\dfrac{\Delta_{AB}}{2}[\cos{(kx)}+\cos{(ky)}]$.

As mentioned earlier, here we need to do a two step canonical transformation in order to diagonalize the Hamiltonian. The first set of transformations are the same as mentioned in section III. We neglect the interband pairing terms from the Hamiltonian obtained after the first set of transformations and perform a regular two band Bogoluibov transformation which is given by
\begin{equation}
\begin{aligned}
   d_{k1\uparrow}=u_{k1}f_{1k}+v_{k1}f_{2k}^{\dagger}\\
   d_{-k1\downarrow}^{\dagger}=-v_{k1}f_{1k}+u_{k1}f_{2k}^{\dagger}\\
    d_{k2\uparrow}=u_{k2}f_{3k}+v_{k2}f_{4k}^{\dagger}\\
   d_{-k2\downarrow}^{\dagger}=-v_{k2}f_{3k}+u_{k2}f_{4k}^{\dagger}\\
\end{aligned}     
\end{equation}

Here, $u_{k1}^{2}=v_{k2}^{2}=\dfrac{1}{2}\bigg(1+\dfrac{\omega_{\uparrow}+\omega_{\downarrow}}{\sqrt{(\omega_{\uparrow}+\omega_{\downarrow})^2+4\nu^2}}\bigg)$ and $u_{k2}^{2}=v_{k1}^{2}=\dfrac{1}{2}\bigg(1-\dfrac{\omega_{\uparrow}+\omega_{\downarrow}}{\sqrt{(\omega_{\uparrow}+\omega_{\downarrow})^2+4\nu^2}}\bigg)$ where,
$\omega_{\sigma}=h_{1\sigma}(k)(\alpha_{k\sigma}^{2}-\beta_{k\sigma}^{2})-2h_{2\sigma}(k)\alpha_{k\sigma}\beta_{k\sigma}$
and $\nu=-h_3(k)(\alpha_{k\uparrow}\beta_{k\downarrow}+\alpha_{k\downarrow}\beta_{k\uparrow})$.

The self-consistent equations for various order-parameters are given below.

\begin{align}
    \Delta_{AB}=&\langle c_{iA\uparrow}^{\dagger}c_{jB\downarrow}^{\dagger}\rangle-\langle c_{iA\downarrow}^{\dagger}c_{jB\uparrow}^{\dagger}\rangle \nonumber \\
=\dfrac{1}{N}\sum_{k}&(\alpha_{k\downarrow}\beta_{k\uparrow}u_{k2}v_{k2}-\alpha_{k\uparrow}\beta_{k\downarrow}u_{k1}v_{k1})\gamma_{sc}(k)
    \end{align}
with $\gamma_{sc}(k)=\cos{(k_{x})} \pm \cos{(k_{y})}$. The plus sign is for the extended s-wave symmetry while the  minus sign is for the d-wave symmetry in the pairing amplitude.

The magnetization on the A sublattice is equal and opposite to the magnetization on the B sublattice owing to particle-hole symmetry of the Hamiltonian at half-filling. Hence the staggered magnetization $m_s=(m_A-m_B)/2=m_A$.
\begin{align}
m_s=&\langle n_{A\uparrow}- n_{A\downarrow}\rangle\nonumber \\=&\dfrac{1}{N}\sum_{k}[(\alpha_{k\uparrow}^2-\alpha_{k\downarrow}^2)v_{k1}^2+(\beta_{k\uparrow}^2-\beta_{k\downarrow}^2)v_{k2}^{2}]
\end{align}
The density difference between A and B sublattices, also equal to the doublon density on the A sublattice and the hole density on the B sublattice, is given by
\begin{align}
\delta=&\dfrac{\langle n_{A}\rangle-\langle n_{B}\rangle}{2}\nonumber \\=&\dfrac{1}{2N}\sum_{k\sigma}[\alpha_{k\sigma}^2(v_{k1}^2-v_{k2}^2)+\beta_{k\sigma}^2(v_{k2}^{2}-v_{k1}^{2})]
     \end{align}
 $\chi_{AB\sigma}$, defined below, gives the inter-sublattice Fock-shift which comes from the mean field decomposition of the exchange term and the trimer terms in the low energy effective Hamiltonian in Eq.~\ref{H_ren}. 
\begin{align}
      \chi_{AB\sigma}&=\langle c_{iA\sigma}^{\dagger}c_{jB\sigma} \rangle \nonumber\\
     &=\dfrac{1}{4N}\sum_{k}\alpha_{k\sigma}\beta_{k\sigma}(v_{k2}^{2}-v_{k1}^{2})\gamma_{k}
     \end{align}
Similarly, $\chi_{BB\sigma}$ and $\chi_{BBxy\sigma}$ represent second neighbour hoppings within the B sublattice obtained by the mean-field decomposition of the trimer terms and are given by
    \begin{align}
      \chi_{BB\sigma}&=\langle c_{iB\sigma}^{\dagger}c_{jB\sigma}+h.c.\rangle \hspace{1cm}\text{j=i$\pm$2x or i$\pm$2y}\nonumber\\
      &=\dfrac{1}{N}\sum_{k}[\cos{2k_{x}}+\cos{2k_{y}}](\alpha_{k\sigma}^2v_{k2}^2+\beta_{k\sigma}^2v_{k1}^2)\\
      &\nonumber\\
      \chi_{BBxy\sigma}&=\langle c_{iB\sigma}^{\dagger}c_{jB\sigma}+h.c.\rangle \hspace{0.5cm}\text{j=i$\pm$ x$\pm$ y}\nonumber\\
    =\dfrac{1}{N}&\sum_{k}2\cos{(k_{x})}\cos{(k_{y})}(\alpha_{k\sigma}^2v_{k2}^2+\beta_{k\sigma}^{2}v_{k1}^{2})
\end{align}
The spin symmetric RMFT can be obtained from the generic equations, described above, by imposing the spin symmetry. 
\\
{\bf Results for Fock Shift}: 
 \begin{figure*}[!htbp]
    \centering
    \includegraphics[scale=0.4,angle=-90]{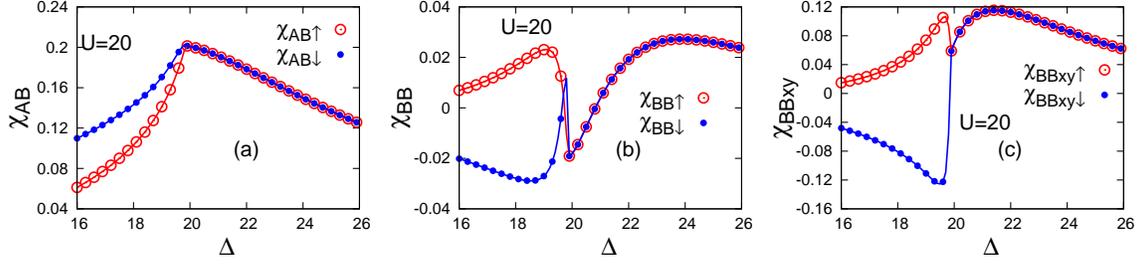}
    \caption{Inter and intra sublattice fock shifts obtained from the generic RMFT which allows for spin symmetry breaking. Panel (a) shows inter sublattice fock shifts $\chi_{AB\sigma}$ vs $\Delta$ while panel (b) shows intra sublattice fock shift $\chi_{BB\sigma}$ along the $2x$ or $2y$ bond. Panel (c) shows intra sublattice fock shift $\chi_{BBxy\sigma}$. Effects due to the phase transitions from the AF-MI to the paramagnetic BI phase [see Fig.~\ref{sc_AF}] are clearly present here as well.}
\label{chis}
  \end{figure*}
 Fig.~\ref{chis} shows the variation of the inter and intra sublattice fock shifts as a function of $\Delta$ for $U=20t$. The inter sublattice fock shift first increases with increase in $\Delta$ with $\chi_{AB\downarrow}>\chi_{AB\uparrow}$, reaches a maximum near the AF transition point, and then decreases with increase in $\Delta$ in the paramagnetic phase. This is because in the AF ordered regime, the density difference between the two sublattices is very near to zero but increases slowly with increasing $\Delta$ due to the presence of some doublons on the A sublattice and holes on the B sublattice. $\chi_{AB}$ in both the spin channels increase due to the increased hopping probability. But beyond the magnetic transition point, densities of doublons on the A sublattice and holes on the B sublattice increase quite rapidly, resulting in an increasing charge density wave insulating behaviour with increasing $\Delta$; hence $\chi_{AB}$ in the paramagnetic regime decreases with increase in $\Delta$. This is shown in panel (a) of Fig.~\ref{chis}. Panel (b) shows the intra sublattice fock shift on the B sublattice, with two B sites separated by next neighbour spacings in either the x or y direction on the square lattice. While $\chi_{BB\uparrow}$ initially increases and then decreases in the magnetically ordered phase,  $\chi_{BB\downarrow}$ decreases and then increases and finally the two become equal to each other in the paramagnetic phase. Panel (c) shows the behaviour of $\chi_{BBxy}$ with $\Delta$ which is the B sublattice fock shift for the two B sites separated by one unit spacing along the x direction and one unit spacing along the y direction. It shows a behaviour qualitatively similar to $\chi_{BB}$.
\section{Appendix B}
In this appendix, we provide details of the full numerical diagonalization of the mean field Hamiltonian. We also show a comparison of the results of this calculation with our earlier calculations where inter-band terms were ignored. The comparison shows that the inter-band terms have a very weak effect on all physical quantities of interest at zero temperature. Thus the phase-diagram we have obtained remains same both qualitatively and quantitatively even in this full numerical calculation. In the following discussion, we will refer to these calculations as the calculation with inter-band pairing terms and without the inter-band pairing terms.

We diagonalise the mean field Hamiltonian by a transformation\\

$\left({\begin{array}{c}
     c_{kA\uparrow}\\
     c_{kB\uparrow}\\
     c_{-kA\downarrow}^{\dagger}\\
     c_{-kB\downarrow}^{\dagger}
     \end{array}}\right)=\left({\begin{array}{cccc}
        u_{1k\uparrow}  & u_{2k\uparrow} & v_{1k\uparrow} & v_{2k\uparrow} \\
        u_{3k\uparrow}  & u_{4k\uparrow} & v_{3k\uparrow} & v_{4k\uparrow} \\ 
        -v_{1k\downarrow}  & -v_{2k\downarrow} & u_{1k\downarrow} & u_{2k\downarrow} \\
         -v_{3k\downarrow}  & -v_{4k\downarrow} & u_{3k\downarrow} & u_{4k\downarrow} \end{array}}\right)\left({\begin{array}{c}
     f_{1k}\\
     f_{3k}\\
     f_{2k}^{\dagger}\\
     f_{4k}^{\dagger}
     \end{array}}\right)$
     \\

 After the transformation, the diagonalized Hamiltonian is assumed to have the form $\mathcal{H}=\sum_{k,\alpha}E_{\alpha}(k)f_{\alpha k}^{\dagger}f_{\alpha k}+const$. We calculate the commutators of the fermionic  $c_{k A,B}$ operators with the mean field Hamiltonian and the diagonalized Hamiltonian and equate the coefficients of the Bogoluibov operators $f_{ik}$ for $i=1,4$ to obtain the eigenvalue equations. Finaly we solve the eigenvalue equation numerically for every k-value in the Brillouin zone to get the eigenvectors and obtain various physical quantities using the following self-consistent equations.   
  
      \bea
          \chi_{AB\sigma}=&\dfrac{1}{4N}\sum_{k}(v_{1k\sigma}v_{3k\sigma}+v_{2k\sigma}v_{4k\sigma})\gamma_{k} \nonumber \\
          \chi_{BB\sigma}=&\dfrac{1}{N}\sum_{k}(v_{3k\sigma}^{2}+v_{4k\sigma}^{2})(\cos{(2k_{x})}+\cos{(2k_{y})})\nonumber \\
          \chi_{BBxy\sigma}=&\dfrac{1}{N}\sum_{k}(v_{3k\sigma}^{2}+v_{4k\sigma}^{2})(2\cos{(k_{x})}\cos{(k_{y})})\nonumber \\
          \delta=&\dfrac{1}{2N}\sum_{k,\sigma}(v_{1k\sigma}^{2}+v_{2k\sigma}^{2}-v_{3k\sigma}^{2}-v_{4k\sigma}^{2}) \nonumber \\
          m_s=&\dfrac{1}{N}\sum_{k}(v_{1k\uparrow}^{2}-v_{1k\downarrow}^{2}+v_{2k\uparrow}^{2}-v_{2k\downarrow}^{2}) \nonumber \\
          \Delta_{d,s}=&\dfrac{1}{N}\sum_{k}(v_{1k\uparrow}u_{3k\downarrow}+v_{2k\uparrow}u_{4k\downarrow})\gamma_{sc}(k)\nonumber 
      \eea
     
    \begin{figure}[!htbp]
    \centering
    \includegraphics[width=5cm,angle=-90]{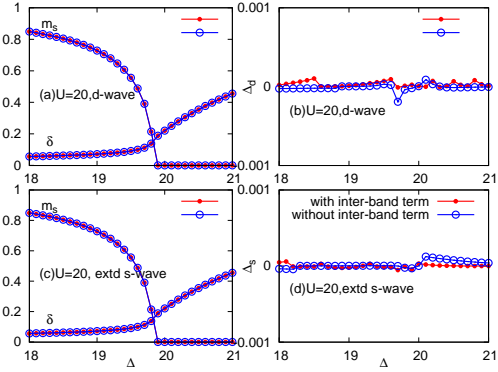}
    \caption{The staggered magnetization $m_s$ and the density difference $\delta$ as functions of $\Delta$ for $U=20t$. The top left panel shows the data for d-wave pairing and the bottom left panel for the extended s-wave case. Right panels show the pairing amplitudes for the d-wave and extended s-wave pairing for $U=20$. As shown, the effect of including inter band pairing in the spin-asymmetric case is negigible.}
\label{ms_comp}
  \end{figure} 
\vskip0.5cm
{\bf{Comparison of results}}:  
We first compare the results of the two calculations with and without inter-band pairing terms for the case where magnetic order is allowed along with the SC order. As shown in Fig.~\ref{ms_comp}, the staggered magnetization and the density difference in the two calculations are exactly the same. The pairing amplitudes for the d-wave and the extended-s wave pairing are shown in right panels of Fig.~\ref{ms_comp}. Superconductivity does not turn on even in the calculation with inter-band pairing and the pairing amplitude for both the d-wave and the extended s-wave symmetry remains zero. 

  \begin{figure}
    \centering
    \includegraphics[width=3cm,angle=-90]{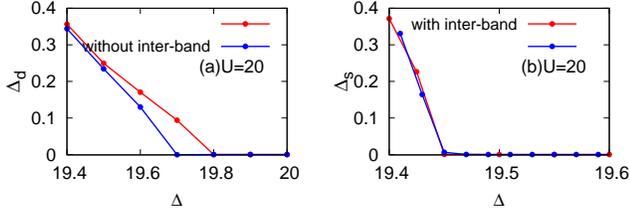}
    \caption{The left panel shows the d-wave pairing amplitude for $U=20t$ in the spin-symmetric calculation. There is a small change in the d-wave pairing amplitude due to the inter-band pairing terms which lead to a small enhancement of the pairing amplitude. The right panel shows the pairing amplitude for the extended s-wave symmetry. Inter band pairing terms have an even weaker effect on the extended s-wave pairing amplitude than on the d-wave pairing amplitude.}
\label{d_comp}
  \end{figure}
We have also compared the results for the case where the spin symmetry is enforced and only the SC order is allowed. In this case, the transformation used to diagonalise the mean field Hamiltonian gets simplified due to the smaller number of variables involved. Here, due to spin symmetry $v_{ik\ua}=v_{ik\da}$ and $u_{ik\ua}=u_{ik\da}$ for $i=1,4$.
Fig.~\ref{d_comp} shows the d-wave pairing amplitude as a function of $\Delta$ for the calculations with and without inter-band pairing terms.  There is a weak effect of the inter-band pairing term on the d-wave pairing amplitude though the range in $\Delta$ over which $\Delta_d$ remains non-zero is more or less same in the two calculations. The effect of the inter-band pairing on the extended s-wave pairing amplitude is even weaker as shown in the right panel of Fig.~\ref{d_comp}.   

\end{document}